\documentclass[11pt,english,nofootinbib,onecolumn]{revtex4-1}

\usepackage[font=small,labelfont=bf]{caption}
\usepackage{subcaption}
\usepackage{wrapfig}
\usepackage{float}
\floatstyle{boxed}
\restylefloat{figure}
\usepackage[pdftex]{graphicx}
\usepackage{savesym}
\usepackage{listings}
\usepackage{float}
\usepackage{epsfig}
\restylefloat{figure}
\floatstyle{boxed}
\usepackage{amsthm}
\usepackage{babel}
\usepackage[intlimits]{amsmath}
\savesymbol{iint}
\usepackage{txfonts}
\restoresymbol{TXF}{iint}
\usepackage{amssymb}
\usepackage{comment}
\usepackage{color}
\usepackage{cancel}

\def\be{\begin{equation}}
\def\ee{\end{equation}}
\def\ba{\begin{eqnarray}}
\def\ea{\end{eqnarray}}
\def\go{\mathrel{\raise.3ex\hbox{$>$}\mkern-14mu
             \lower0.6ex\hbox{$\sim$}}}
\def\lo{\mathrel{\raise.3ex\hbox{$<$}\mkern-14mu
             \lower0.6ex\hbox{$\sim$}}}

\usepackage{babel}
\begin{document}

\title{The Blue Box White Paper}
\author{Doctor Benjamin K. Tippett\footnote{email: barn@titaniumphysics.com}}
\affiliation{Gallifrey Polytechnic Institute }

\author{Doctor David Tsang\footnote{email: dtsang@physics.mcgill.ca}}
\affiliation{Gallifrey Institute of Technology (GalTech) }

\date{\today}
\begin{abstract}
This white paper is an explanation of Ben and Dave's TARDIS time machine, written for laypeople who are interested in time travel, but have no technical knowledge of Einstein's Theory of General Relativity. 

The first part of this paper is an introduction to the pertinent ideas from Einstein's theory of curved spacetime,  followed by a review of other popular time machine spacetimes. We begin with an introduction to curvature and lightcones. We then  explain the Alcubierre Warp Drive, the Morris-Thorne wormhole, and  the Tipler cylinder.
 
We then describe the Traversable Achronal Retrograde Domain in Spacetime (TARDIS), and explain  some of its general properties. Our TARDIS is a bubble of spacetime curvature which travels along a closed loop in space and time. A person travelling within the bubble will feel a constant acceleration. A person outside of the TARDIS will see two bubbles: one which is evolving forwards in time, and one which is evolving backwards in time. We then discuss the physical limitations which may prevent us from ever constructing a TARDIS.

Finally, we discuss the method through which a TARDIS can be used to travel between arbitrary points in space and time, and the possible dangers involved with exiting a TARDIS from the wrong side.

Before we begin, would you like a Jelly Baby?
  \end{abstract}
\maketitle
\part*{Part 1: Background}
\section*{General Relativity}
\subsection*{Einstein's Theory of Curved Spacetime }

\begin{quote}
Spacetime tells matter how to move; matter tells spacetime how to curve.

-John Archibald Wheeler
\end{quote}

Let us begin by introducing Albert Einstein's greatest achievement\footnote{Seriously, have a Jelly Baby. They're great! Or a Banana. Bananas are fantastic!}: the Theory of General Relativity. According to this theory, the three spatial dimensions are not formally separate from the dimension of time. Rather, the four directions (left-right, up-down, forward-backward, and future-past) comprise a four dimensional surface, across which the stars and planets and all the rest of the matter in the universe tumble. We interpret the effects of this curvature as the force of gravity.

\begin{figure}
\includegraphics[trim=10mm 1mm 1mm 11mm, clip, width=11cm]{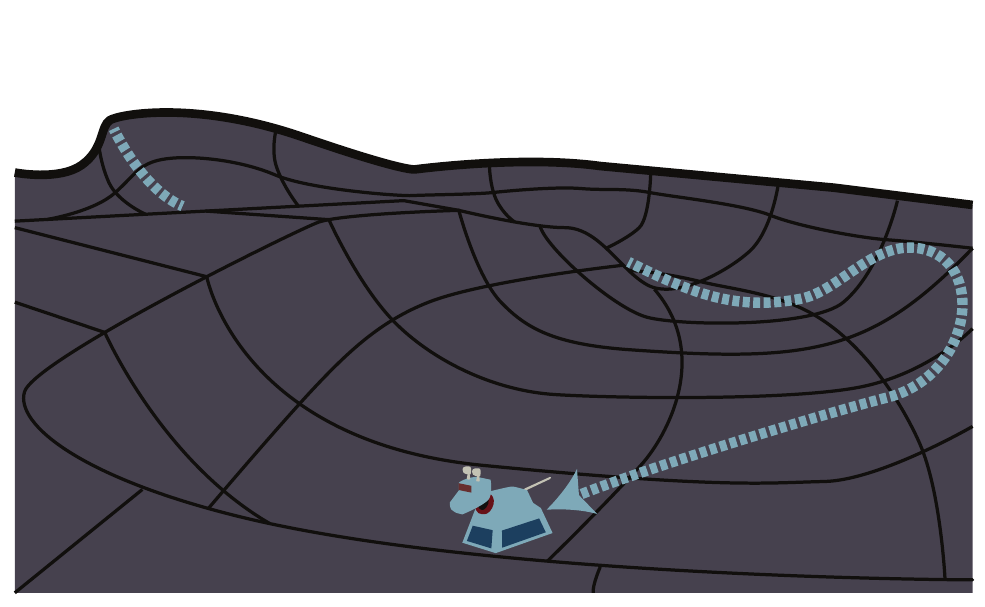}

\caption{We ask K9 to steer straight-ahead as he moves over the curved ground of an old quarry in Wales. The curvature of the surface is manifested by causing the tin dog's path to bend and skew.  \label{K9}}

\end{figure}

Imagine an old quarry, the floor of which is covered in smooth lips and bumps and bowls. Suppose that one were to ask one's robot dog to enter the quarry, and to fix its steering wheels in the forward pointing position. Over flat ground, if the instructions were followed, the tin dog would move along a straight line. The curved terrain of the quarry will cause the companion's path to bend and twist and skew (Fig. \ref{K9}).

Similarly, in Einstein's theory the curvature of spacetime accounts for the curved orbits of the planets. In ``flat" (uncurved) spacetime, planets and stars would move along straight lines. In the vicinity of a massive star the spacetime geometry becomes curved, and the trajectories of nearby planets will bend around star. 

Unlike the geometry of the floor of the quarry, in Einstein's theory the curvature is dynamic. Both the degree to which the spacetime is curved and the character of the curvature  depend upon the quantity and character of the matter present. The mathematical relationship between the curvature and the matter is called the \emph{Einstein Equation}.

There are two different strategies which are used to ``solve"  the Einstein equation.

We could begin by focussing on the  matter in the spacetime.  A physicist could choose a type of matter and an initial 3-dimensional configuration, and then will use a computer to simulate the evolution of  the geometry and the matter. The computationally generated history of the 3-dimensional system then comprises the dynamical four dimensional spacetime.

\begin{figure}
\includegraphics[trim=0mm 0mm 0mm 0mm, clip, width=11cm]{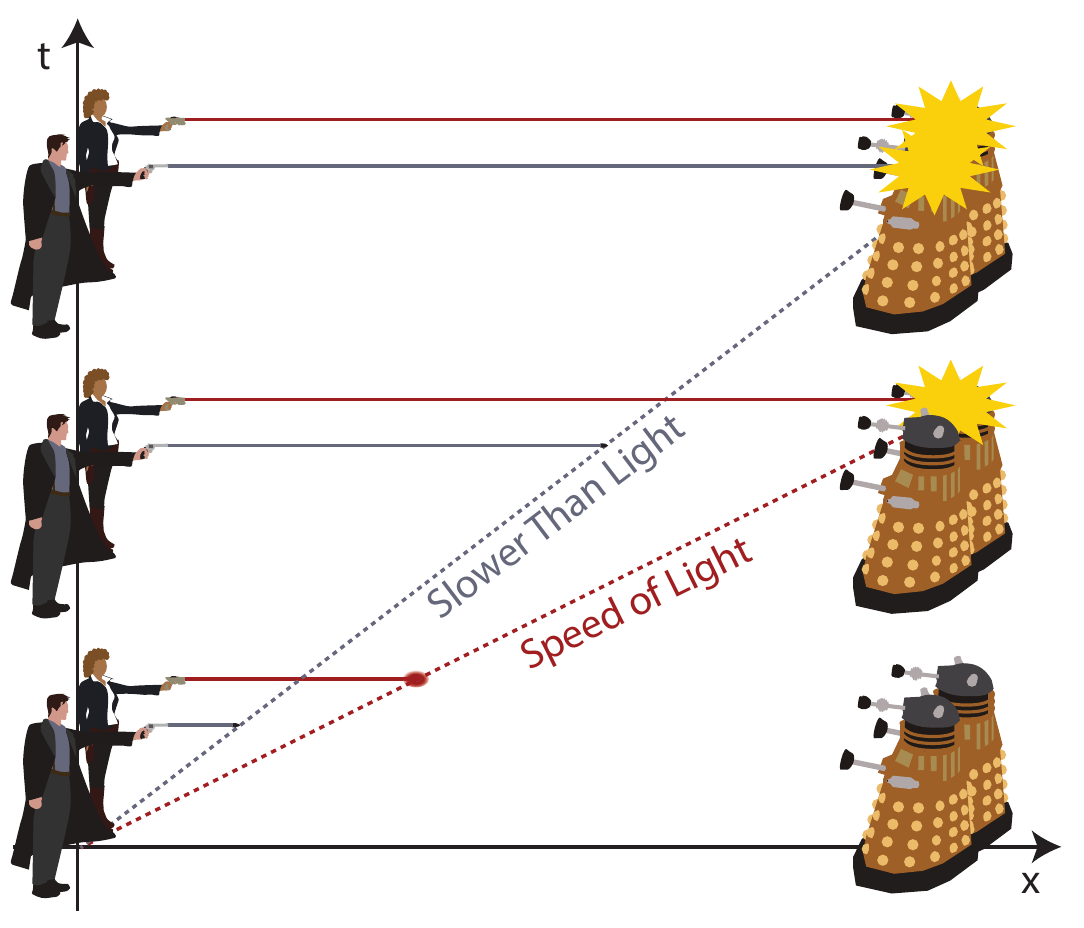}

\caption{Nothing which possesses mass can move faster than the speed of light.  The two companions pull their triggers at the same time. Captain Jack's lead bullet, no matter how awesome awesome his gun is, will never beat River Song's laser to the Dalek. This is an example of a \emph{spacetime diagram}, with space on the horizontal axis and time on the vertical axis. The dashed lines are the respective trajectories of the bullet and the laser pulse.  The laser pulse has a shallower slope than the bullet because it is moving faster than the bullet.  \label{nulltime}}

\end{figure}

\begin{figure*}
\includegraphics[trim=0mm 0mm 0mm 0mm, clip, width=11cm]{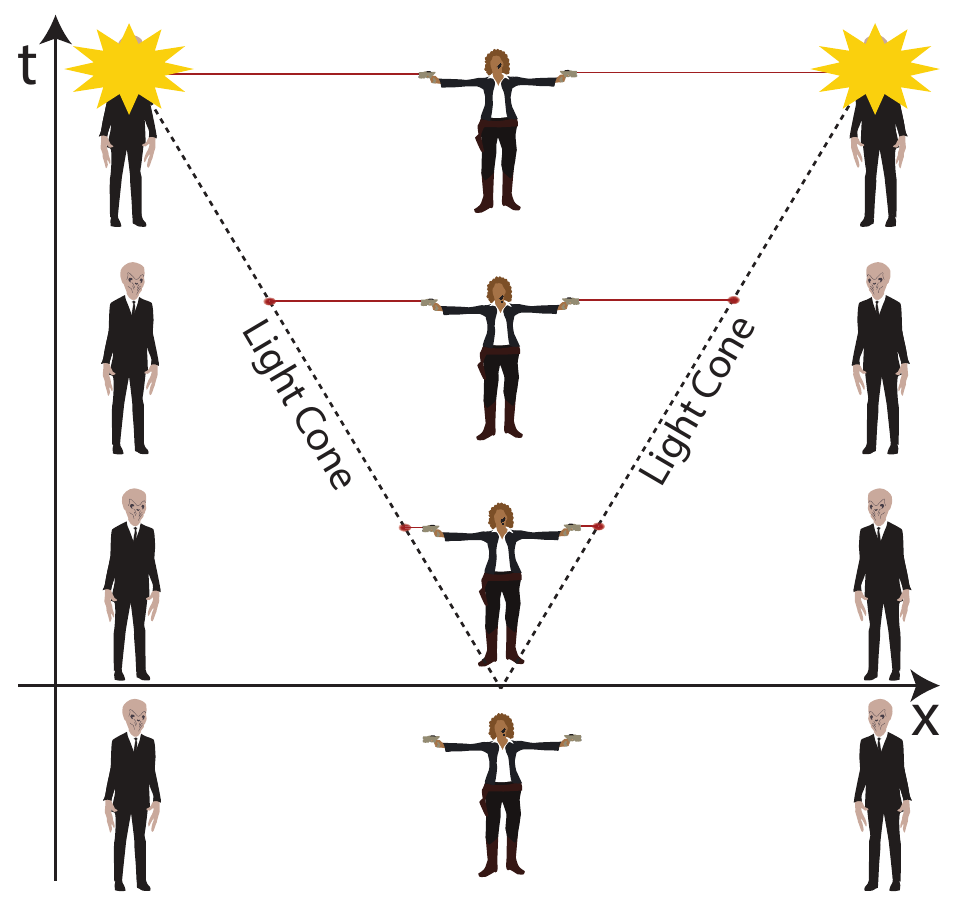}

\caption{Oh, dear! River finds herself facing down two silents, one on either side. Luckily, she is wonderfully adept at using her laser pistols, and immediately shoots a stream of deadly photons in both directions. By plotting the trajectories of photons emitted in all possible directions  from one point, we generate a \emph{lightcone}. A lightcone divides spacetime into two regions. If a point lies within the lightcone,  it can be reached at a velocity slower than the speed of light \emph{starting from} the point when River pulled her triggers. Conversely, If a point lies outside the lightcone, it is inaccessible for a massive object starting from tip of the lightcone.\label{lightcone}}

\end{figure*}

Alternatively, we could focus on  the geometry of the spacetime. A physicist could define the four dimensional geometry in its entirety and then use the Einstein Equation to determine what type and configuration of matter would be required to generate it.  It is important to analyze this matter skeptically, since many geometries  require types of matter which have never been seen in our universe. The most popular criteria for judging physicality are called the \emph{Classical Energy Conditions}. These conditions require that matter  be gravitationally attractive and that matter  may never travel faster than the speed of light. Any matter which does not satisfy the classical energy conditions is referred to as \emph{unphysical} matter or as \emph{exotic} matter.

\subsection*{Causal structure}

\begin{figure*}
\includegraphics[trim=0mm 0mm 0mm 0mm, clip, width=10cm]{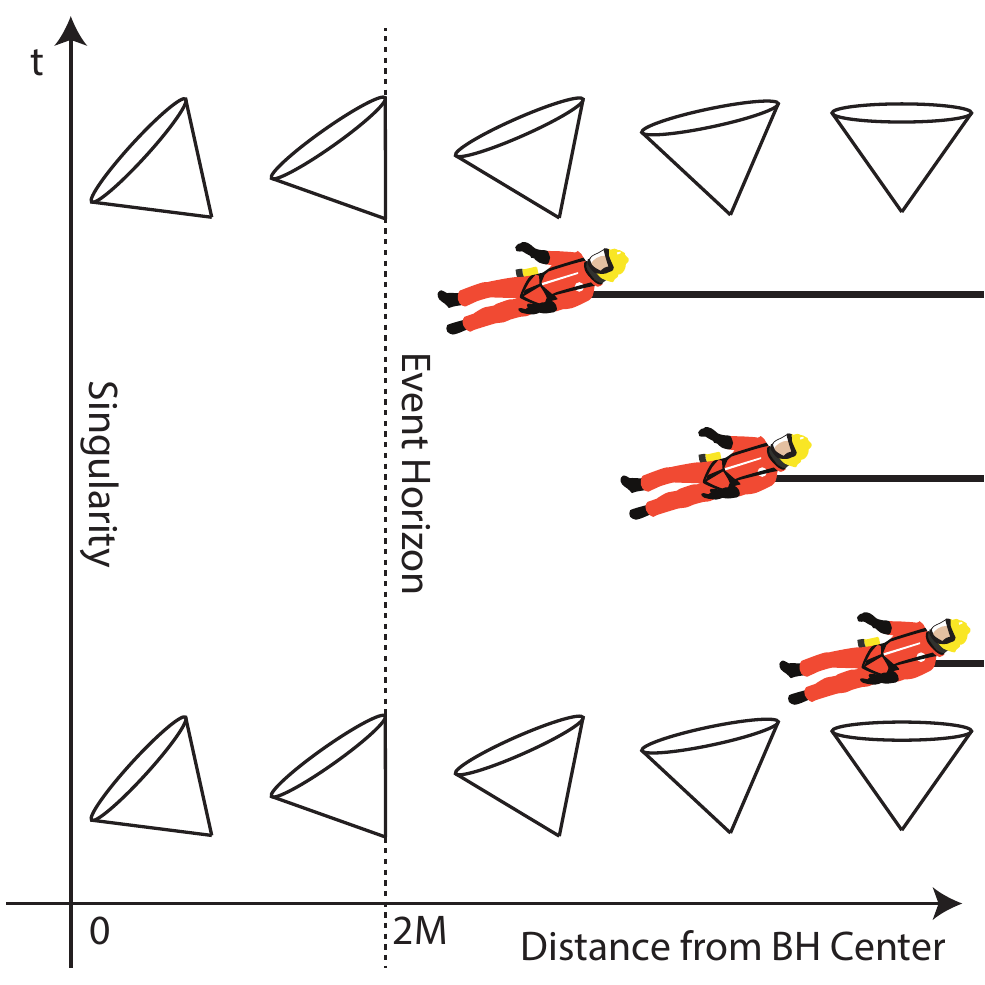}

\caption{Light cones gradually tip inwards, the closer to the centre of a black hole you look. Within the \emph{Event Horizon},  the lightcones tip entirely towards the centre. Thus,  once you have fallen inside a black hole, the orientation of the lightcones will not permit you to climb out  again.   \label{EH}}

\end{figure*}

The fundamental insight underlying Einstein's theory of relativity is that the universe has a speed limit: massive objects must travel slower than the speed of light. In Fig. \ref{nulltime} Captain Jack and River Song are Dalek hunting: Jack is armed with a revolver, and River with a laser pistol. They have come upon a pair of Daleks, and they pull the triggers of their guns simultaneously.  Since Jack's bullets have mass, they must always travel slower than the speed of light, and thus slower than  the photons in River's laser pulse. Therefore, no matter how awesome Jack's gun might be, River's laser pulse will always reach the Dalek ahead of Jack's bullet.  

Fig. \ref{nulltime} is a \emph{spacetime} diagram: it plots how the positions of objects change in time. Since Jack and River are not moving, their trajectories will be a vertical line.  The trajectories of  faster moving objects will have shallower  slopes. Note also that since the laser pulse  travels at the speed of light, no massive object's trajectory may have a shallower slope than the laser's. 

In Figure \ref{lightcone}, River has two lasers, and shoots one pulse to the left and one to the right\footnote{It is ironic that a woman who dislikes silents this much should spend so much time in a library. \emph{SPOILERS!}}. The trajectories of the two laser pulses will trace the \emph{boundary} in spacetime between the points  which can be reached  at subluminal velocities and those which cannot (starting from the point where River pulls the triggers). We call such a boundary a \emph{ lightcone}.

We can draw a lightcone originating from any point in the spacetime, and sometimes it is useful to draw a series of lightcones to illustrate where massive observers are allowed travel. For example, the simplest justification for why a person  cannot ordinarily move \emph{backwards} in time  is because  a looping path across spacetime would cross the lightcone wall, indicating a requirement for superluminal speeds.

Lightcones are incredibly powerful tools for visualizing spacetime curvature. In short, spacetime curvature causes lightcones to \emph{tip}. Thus, mapping out the orientations of the lightcones in a spacetime can illustrate the paths along which a massive object may travel. 

As an example, let us consider a black hole. It is widely known that once you fall into a black hole, you may never emerge.  The  simplest explanation for this fact involves illustrating  the orientations of nearby lightcones  (Fig. \ref{EH}). Near the middle of the black hole the lightcones are tipped entirely towards the central singularity. Thus,  massive objects are corralled by the lightcone walls towards the centre of the black hole.  Alternatively put, escaping a black hole requires faster-than-light speeds.

\section*{Previously Proposed Mechanisms for Time Travel}
\subsection*{The Alcubierre Warp Drive}

\begin{figure}
\includegraphics[trim=70mm 70mm 40mm 65mm, clip, width=10cm]{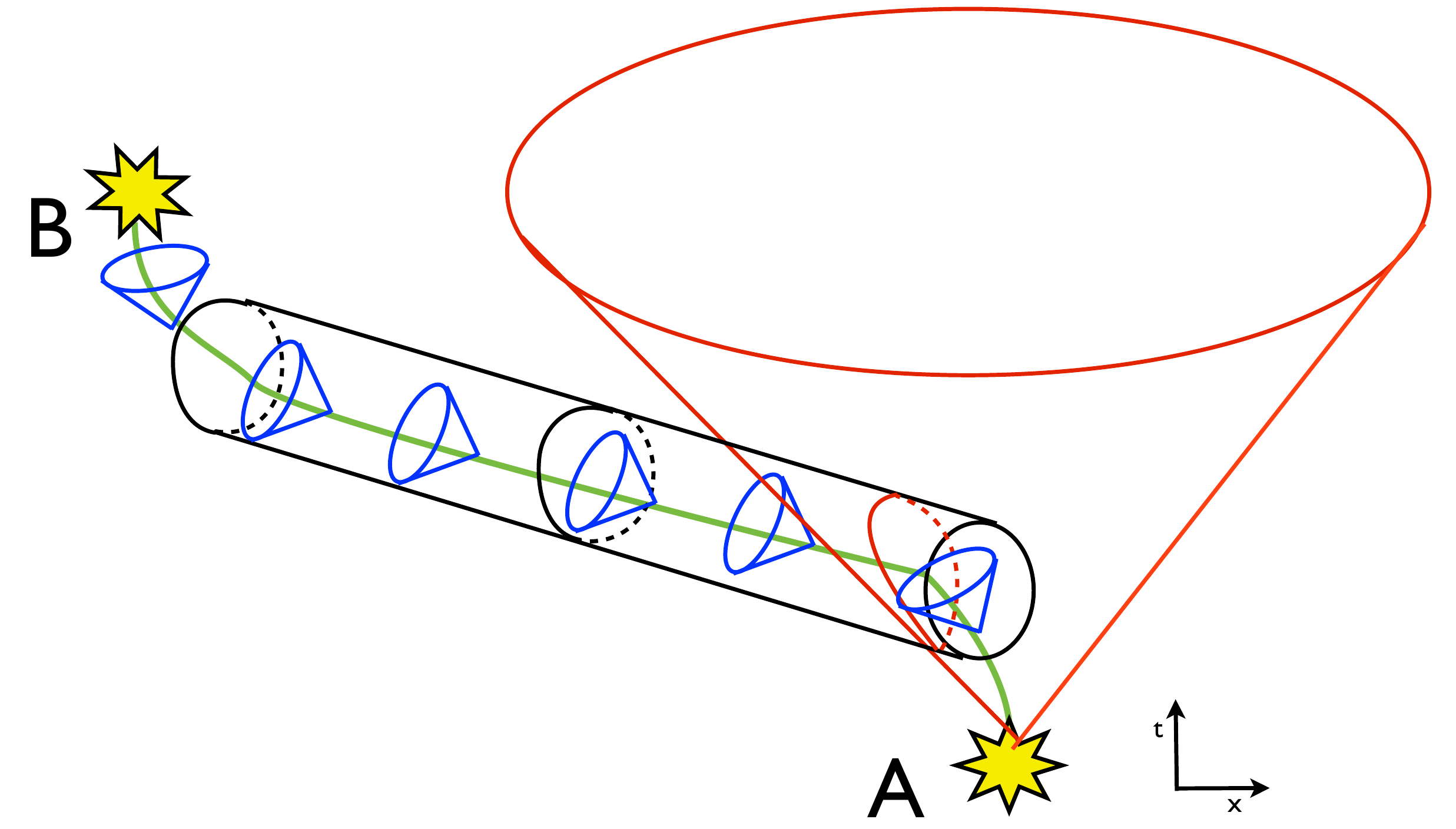}

\caption{The Alcubierre Warp Drive can be used to carry massive objects through spacetime at superluminal velocities. \label{warp}}

\end{figure}

Miguel Alcubierre has proposed a way for a massive object to travel \emph{faster} than the speed of light: the Warp Drive \cite{alcubierre94}. This spacetime geometry can be described as having a bubble configuration: a shell of curvature containing a flat vacuum interior. Spacetime expands and contracts around the edges of the bubble in a manner which allows it to scuttle across the surrounding spacetime at superluminal speeds.  Lightcones contained within the bubble are tilted in whichever direction the bubble is headed, thus massive objects can be carried through spacetime at superluminal speeds (Fig.  \ref{warp}).

As we mentioned, the reason we cannot ordinarily travel backwards in time to interact with our former selves is because doing so requires that we cross the lightcone, and this would require speeds which exceed the speed of light.  Thus, through a judicious choice of a sequence of Alcubierre warp bubbles, a traveller may end up moving backwards through time to interact with their past self \cite{Everett1996}.

How realistic is the Warp Drive? Recall that any geometry in spacetime is accompanied by a specific distribution of matter and mass, which causes it to curve. The matter required to create the Alcubierre Warp Drive geometry violates the classical energy conditions. Thus,  no material which humankind has ever discovered can curve spacetime in the way required to build a Warp Drive.

\subsection*{Wormholes}
\begin{figure}
\includegraphics[trim=57mm 85mm 75mm 15mm, clip, width=7cm]{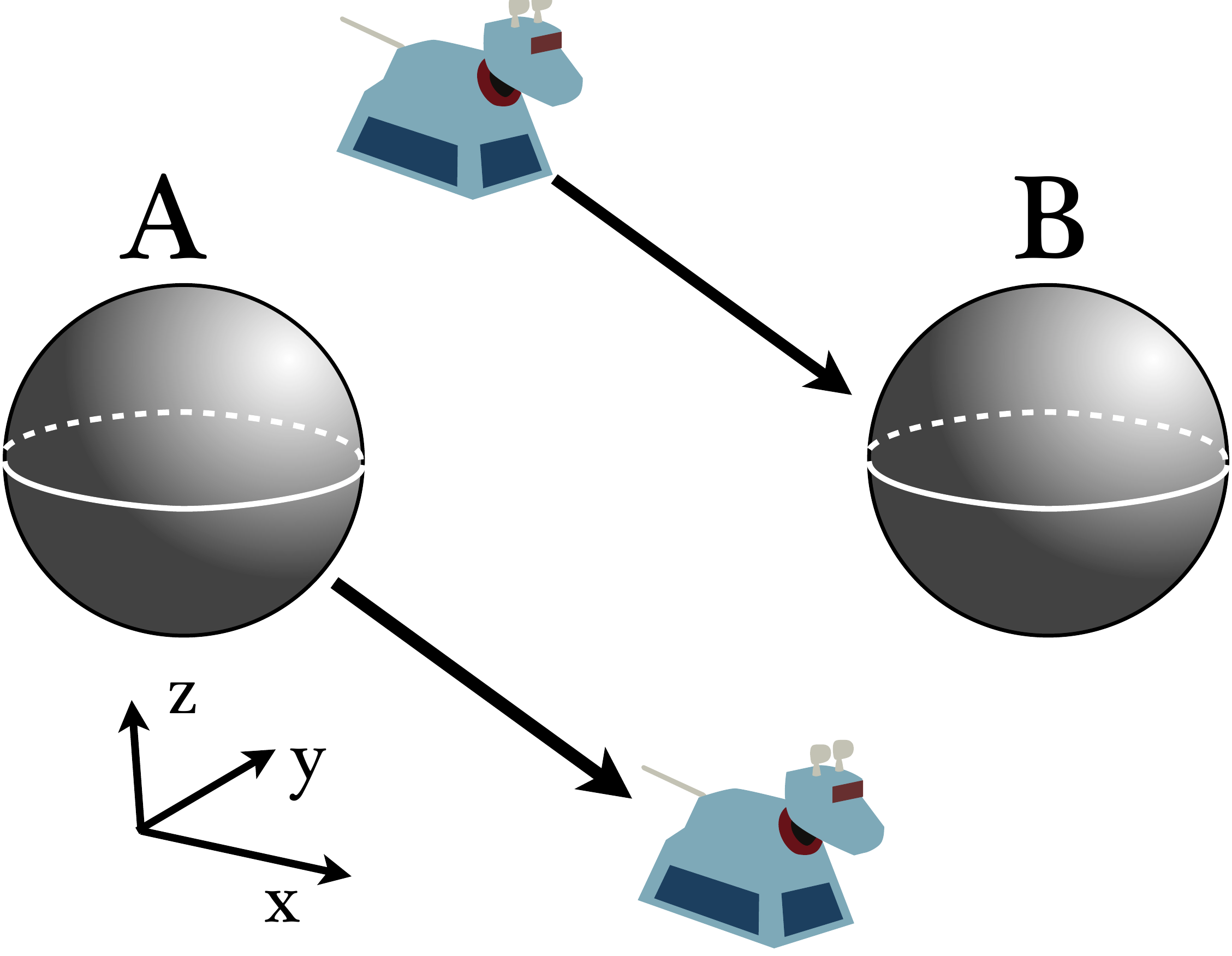}

\caption{A wormhole is drawn as having two mouths (A and B). If your second best friend enters one mouth, it will emerge from the other. \label{TwoM}}

\end{figure}

An alternative way to go backwards in time  was proposed by Michael Morris and Kip Thorne \cite{Morris1988} and takes advantage of an exotic object called a \emph{wormhole}. The simplest way to imagine a wormhole is as a pair of spheres which are  actually the same sphere in two places. If we drive a robot dog into one of the spheres, as it crosses the surface of the first sphere it will  emerge from the second one. We refer to the two spheres as the \emph{mouths} of the wormhole (Fig. \ref{TwoM}).

Suppose we were to take the two wormhole mouths and pack them into boxes, and then give the two boxes to a pair of twins. One of the twins is a space explorer and the other is a gardener. Our first twin sets off in a rocket, visits a distant star,  and then returns to Earth. The story of the twin paradox is familiar: the space twin experiences less time over the course of his journey than his sibling does on earth\footnote{The simple reason for this is that the space twin endures a lot of acceleration in his rocket, and when you accelerate time passes at a slower rate. For a better explanation, listen to The TItanium Physicists Podcast, episode 29.}. Thus, upon their reunion, the space explorer will be younger than his gardener sister.

Let us tell the reunited twins to open their boxes. Just like the twins, the wormhole mouth which journeyed in space will be younger than the mouth which stayed at home. Thus, if you were to throw a baseball at the gardener's mouth, it would emerge from the astronaut's mouth at an earlier point in time.

Morris and Thorne used this type of time machine to perform fascinating thought experiments \cite{Echeverria1991}. What if a billiard ball\footnote{Billiard balls are good for this thought experiment because they are simple and we are under no illusions about free will affecting their behaviour.} travelled back in time only to bump into itself, ultimately preventing a younger version of itself from entering the wormhole in the first place? This series of events would be causally inconsistent with itself. These scenarios are collectively referred to as the \emph{Grandfather Paradox}, referring to the possibility that a time traveler could  kill her grandfather before he has sired her father\footnote{The episode ``Father's Day" is concerned with such a paradox: Rose Tyler  saves her dead father's life\cite{timetravel}.}.

\begin{figure}
\includegraphics[trim=2mm 245mm 95mm 2mm, clip, width=16cm]{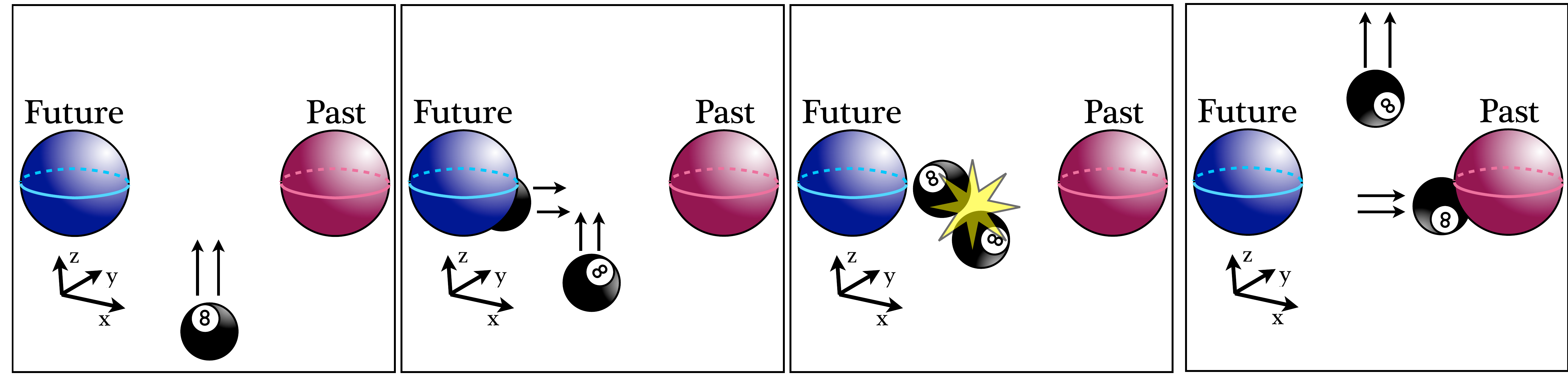}

\caption{A possible way that an object can interact with its past self in a self-consistent way. The two wormhole mouths don't sync up in time, one (red) is older than the other (blue) and can be used to travel into the past. \label{balls}}

\end{figure}

The Grandfather Paradox does not preempt an object from interacting with its past self. Consider a billiard ball which travels backwards in time in order to collide with itself, and knock its past self into the wormhole (Fig. \ref{balls}). Unlike to the previous scenario, this self-interaction would be historically self-consistent. 

Clearly, whenever time travel is possible, the traditional way of describing cause and effect becomes skewed\footnote{Consider, for example, how in ``Smith and Jones," the Tenth Doctor travels backwards in time to take off his necktie at Martha. }. In these cases, we must think of causality in terms of actions and consequences over the entirety of  a four dimensional spacetime: the future can interact with the past so long as the past is not modified. This ``law of nature" is referred to as \emph{ Novikov's self consistency condition} \cite{Friedman1990}. For example, suppose the Doctor were to travel to the planet Skaro and attempt to prevent the creation of the Daleks. Novikov's self consistency condition dictates that all of his efforts would fail \footnote{If you have trouble imagining it, watch ``The Genesis of the Daleks." It has Davros in it. Do you think he can see out of his big blue forehead eyeball? Gross.}, since his motivation for preempting the Daleks will have been\cite{timetravel} predicated upon the Daleks' having existing.

Although Morris and Thorne's model of time travel is  both wonderful and simple,   constructing this geometry is even more difficult than the Warp Drive. One problem is that we must build a worm hole. Physicists are not sure how to even begin punching two  holes in the universe and sewing them together\footnote{We do what we must, because we can.}. Another problem is that the wormhole must be traversable. If any massive object were to attempt to cross through a wormhole, it would cause the wormhole to collapse and pinch off into a pair of black holes. Holding the wormhole open requires a type of unphysical matter which is gravitationally repulsive \cite{Friedman1993}. 

\subsection*{Tipler Cylinder}

\begin{figure}
\includegraphics[trim=0.5mm 0mm 0mm 11mm, clip, width=13cm]{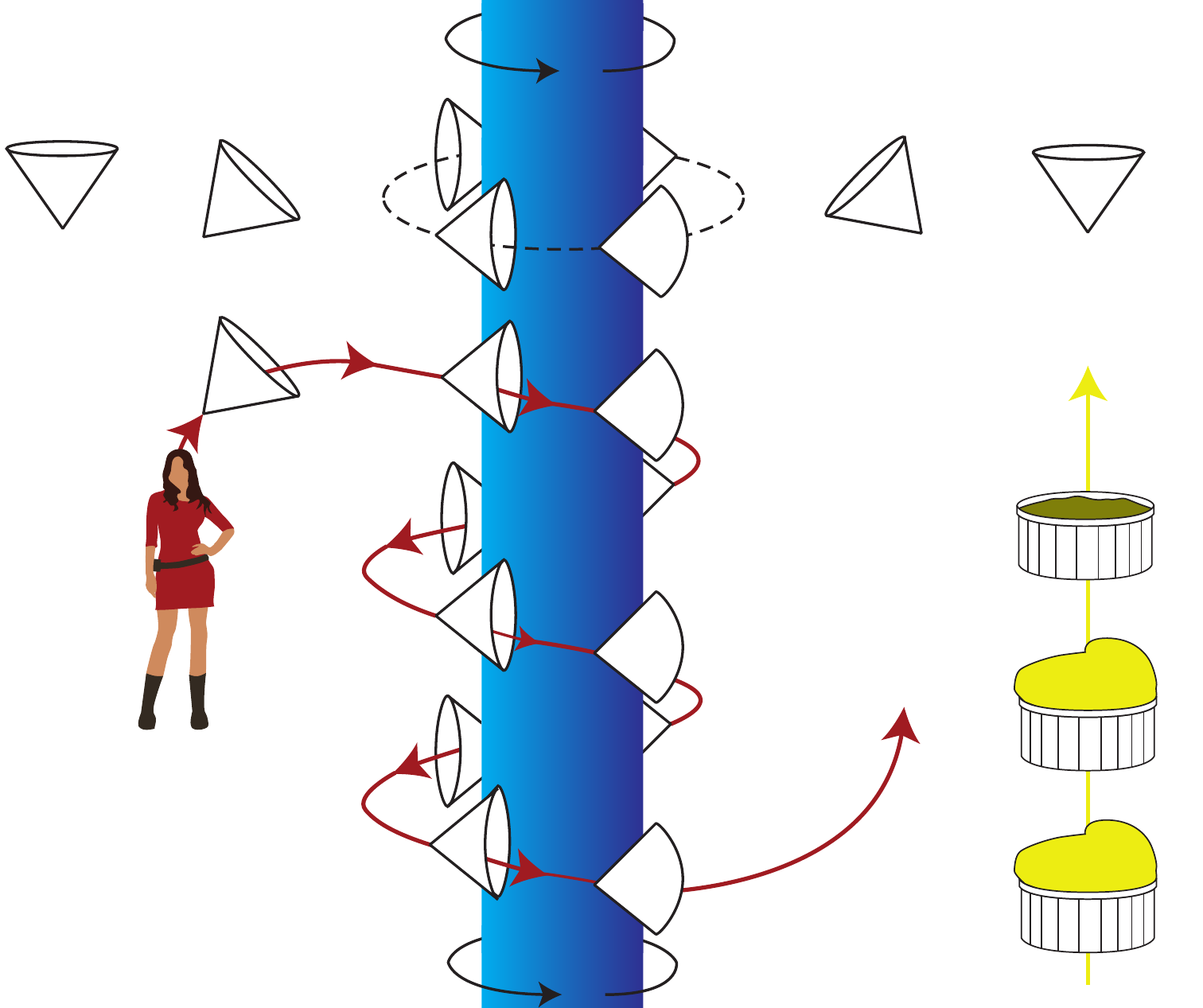}

\caption{Clara discovers that the secret to a fluffy souffl\'{e} is \emph{timing}! Light cones near the edge of the Tipler cylinder tip  towards the direction of rotation. By repeatedly circumnavigating the cylinder, Clara can move into the past and reach her souffl\'{e} before it fell\cite{timetravel2}.  \label{Tiplers}}

\end{figure}

Unphysical and exotic matter is not always required to build a geometry which allows time travel. All you need is an infinite amount of spinning matter! The Tipler cylinder is a very popular method for time travel, requiring only an infinitely long, massive rod, which is spinning on its axis \cite{Tipler1974}. 

Spinning mass has an interesting effect on spacetime: like spinning a spoon in a glass of water, the spacetime swirls around with the object in its the direction of rotation. This effect is called \emph{frame dragging} and it causes lightcones to tip slightly in the direction of the spin. The effect increases with the angular momentum, and with the proximity to the object. The frame dragging effect generated by the Earth is subtle, but it has been detected by satellites \cite{Everitt2011}!

In the case of an infinitely long cylinder, the frame dragging effect becomes quite potent. Near the cylinder surface, the lightcones tip \emph{all the way} over, allowing us to travel backwards in time  by  circling the cylinder (Fig. \ref{Tiplers}).

\part*{Part 2: Explaining the TARDIS}
\section*{the TARDIS geometry}

\begin{figure}
\includegraphics[trim=55mm 80mm 95mm 20mm, clip, width=10cm]{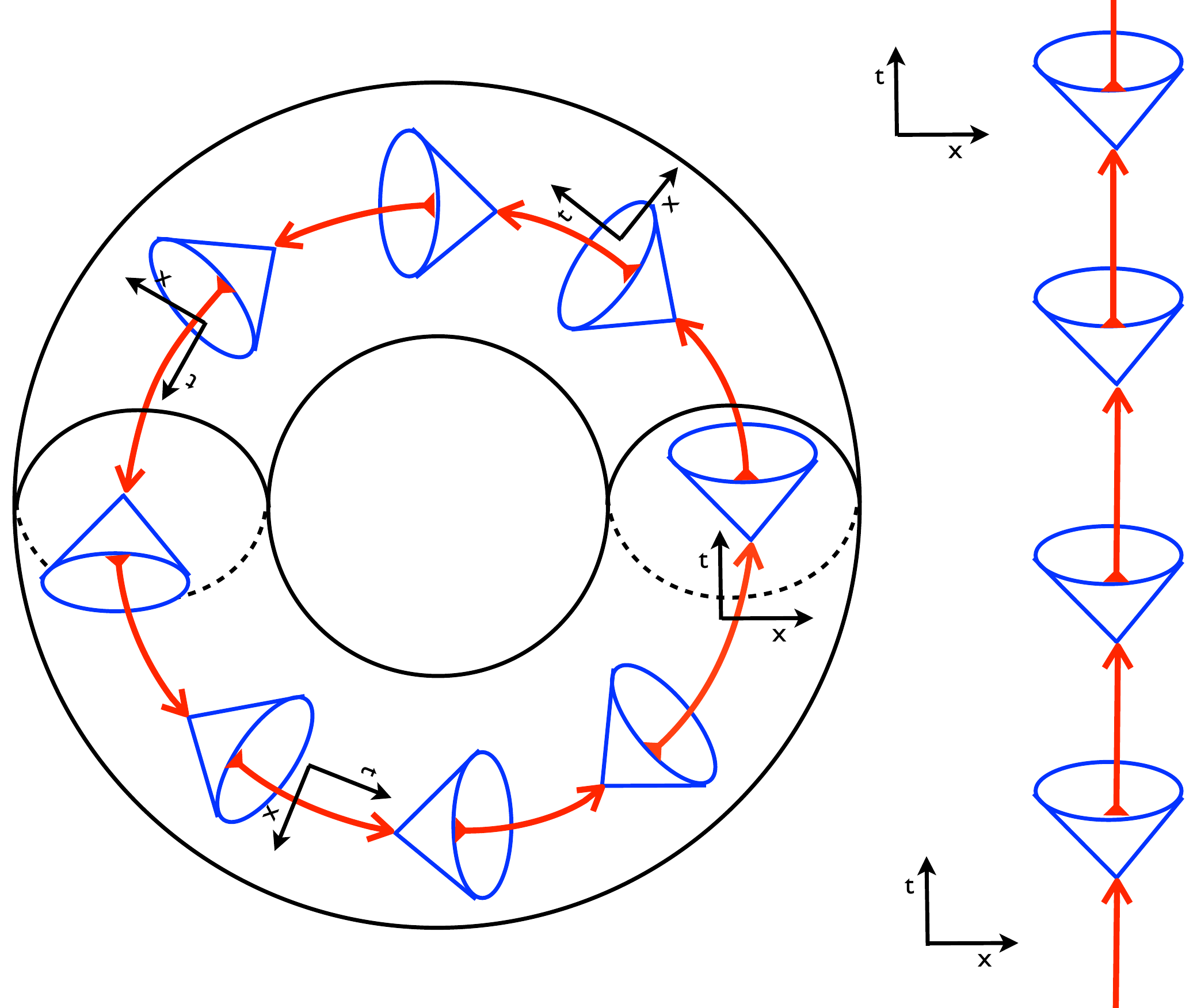}

\caption{This spacetime diagram illustrates how lightcones will be oriented around the TARDIS geometry. Outside of the TARDIS, time is aligned in the  vertical direction. Inside of the bubble, lightcones tip along a circular loop, and massive objects can travel backwards in time. \label{lightconesTAR}}

\end{figure}

\begin{figure}
\includegraphics[trim=6mm 45mm 12mm 10mm, clip, width=12cm]{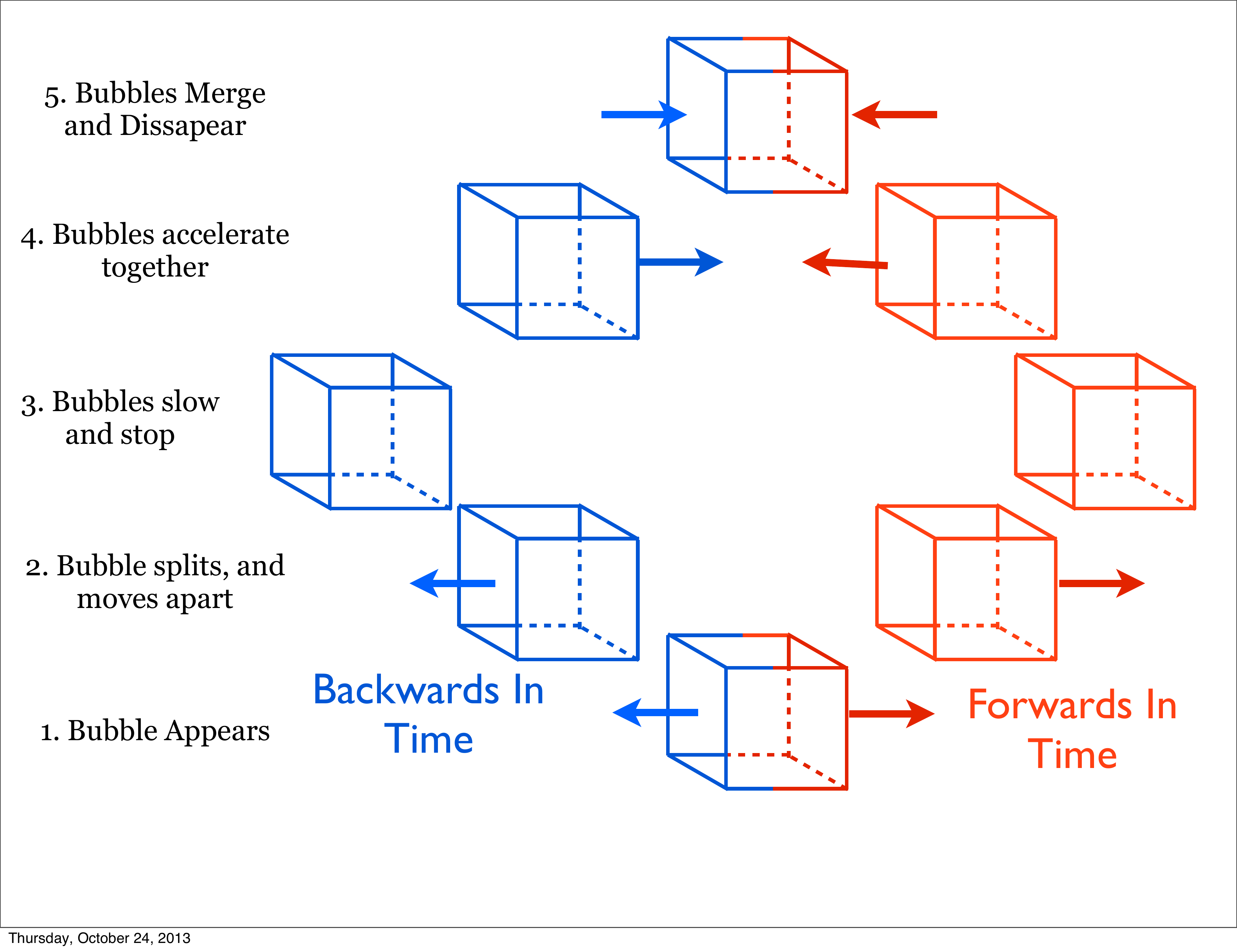}

\caption{An external observer will see two bubbles suddenly emerge from one another, one whose contents appear to move backwards in time. \label{fromTheOUtside}}
\end{figure}

Our TARDIS geometry can be described as a bubble of spacetime geometry which carries its contents backwards and forwards through space and time as it endlessly tours a large circular path in spacetime\footnote{The geometry has a spacetime line-element: $ ds^2=  \left[ 1 -     h(x,y,z,t)   \left( \frac{2t^2}{x^2 +t^2} \right)  \right] (-dt^2 +dx^2)  + h(x,y,z,t)  \left( \frac{4xt}{x^2 +t^2 } \right)dxdt +dy^2+dz^2 $. For more specific information, read the original paper.}. The inside and outside of the bubble are a flat vacuum, and the two regions are separated with a boundary of spacetime curvature.

In this case TARDIS stands for Traversable Achronal Retrograde Domain In Spacetime. The name refers to a bubble (a \emph{Domain})  which moves through the spacetime at speeds greater than the speed of light (it is \emph{Achronal}); it moves backwards in time (\emph{Retrograde} to the arrow of time outside the bubble);   and finally, it can transport massive objects  (it is \emph{Traversable})\footnote{You might be asking ``Why aren't you concentrating on making it bigger on the inside?" The short answer is that it is \emph{very easy} to make a spacetime which is curved so that a very large volume sits inside a very small box. You can even use curved spacetime geometries to make very large objects appear very very small \cite{Tippett2011}! Long story short: Bigger on the inside is too easy to bother.}.

Fig. \ref{lightconesTAR} is a spacetime diagram which shows the orientation of lightcones inside and outside of the  TARDIS bubble. Outside of the bubble, the lightcones are all oriented upwards, pushing everyone steadily towards the future. Inside the bubble, lightcones tip and turn, allowing massive objects to move along a closed circle in spacetime. If a person were to be transported within the bubble,  she would be moving alternatively forwards, sideways and even backwards in time!  

A person travelling within the TARDIS would describe it as a room which is constantly accelerating forwards. Furthermore, any events which occur inside of the TARDIS bubble must satisfy Novikov's self consistency condition\footnote{the Doctor and Romana were stuck in a loop of repeating events in the ``Megalos." The episode also involved a cactus as the villain, so...}.

A person outside the bubble would describe an entierly different scene (see Fig. \ref{fromTheOUtside}). Due to its closed trajectory, there will be a time \emph{before} the bubble (and its contents) exists. Abruptly, we would see a bubble appear and split into two, the two boxes moving away from one another. Initially, they will be moving at  superluminal speeds, but they will decelerate to a stand-still. The pair of boxes will then begin accelerating towards one another until they exceed the speed of light,  merge and disappear. Mysteriously, the contents of one bubble will be appear to evolve forwards in time, while the other will evolve backwards.

\begin{figure}
\includegraphics[trim=0mm 0mm 0mm 0mm, clip, width=10cm]{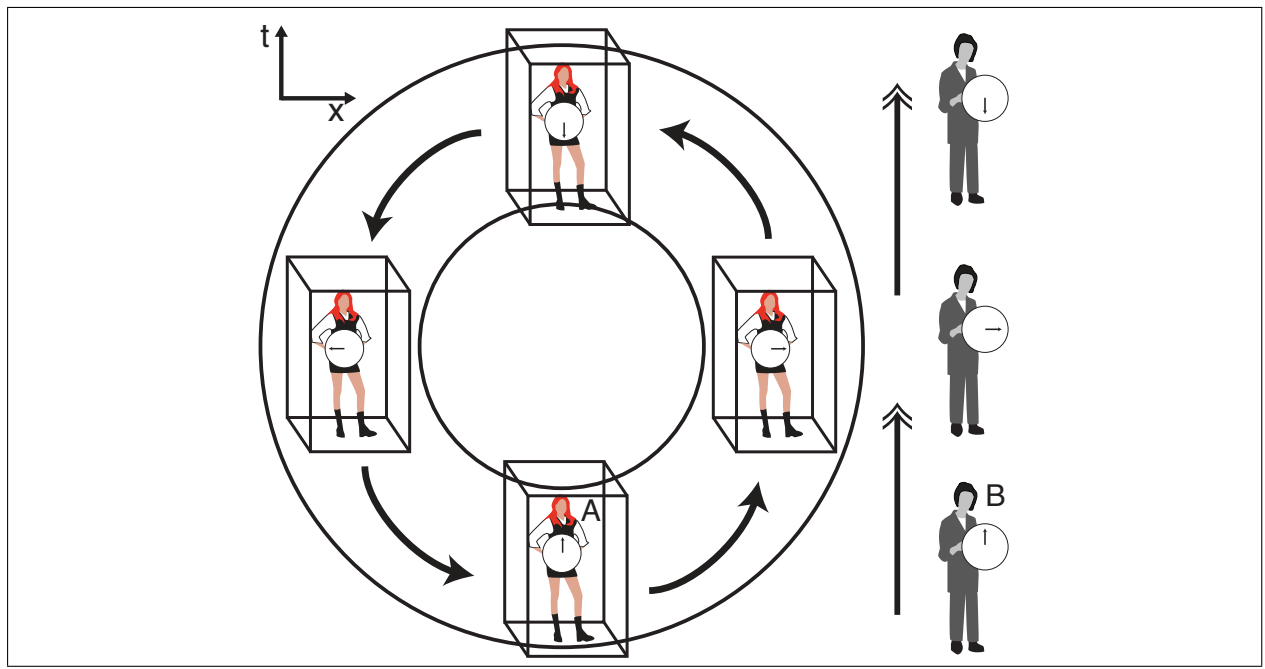}

\caption{This is a  schematic of the trajectory of the TARDIS bubble as it travels through spacetime. Arrows denote the locally defined ``arrow of time."  Life inside the bubble is colourful and sexy and fun. Life outside the bubble is grey and drab, and everyone dresses like a school teacher from the 1960's.  \label{clocks}}

\end{figure}

To illustrate  how time is experienced inside and outside of the bubble, imagine that there are two people in our spacetime: Amy, who is travelling inside the bubble; and Barbara, who has been left behind. Suppose that the two women are holding large clocks, and that the walls of the bubble are transparent, so that the two women can see one another (See Fig. \ref{clocks}).

Amy will only ever see the hands of her own clock move in a clockwise direction. When she looks out at Barbara, she will see the hands of Barbara's clock moving clockwise  at some times and counterclockwise at others, depending on where Amy is along her circular trajectory.

Barbara will see the hands of her own clock moving clockwise. She will see two bubbles, each one containing an Amy. In one of them, the hands of Amy's clock will be moving clockwise; in the other, counterclockwise.

\begin{figure*}
\includegraphics[trim=15mm 25mm 15mm 5mm, clip, width=10cm]{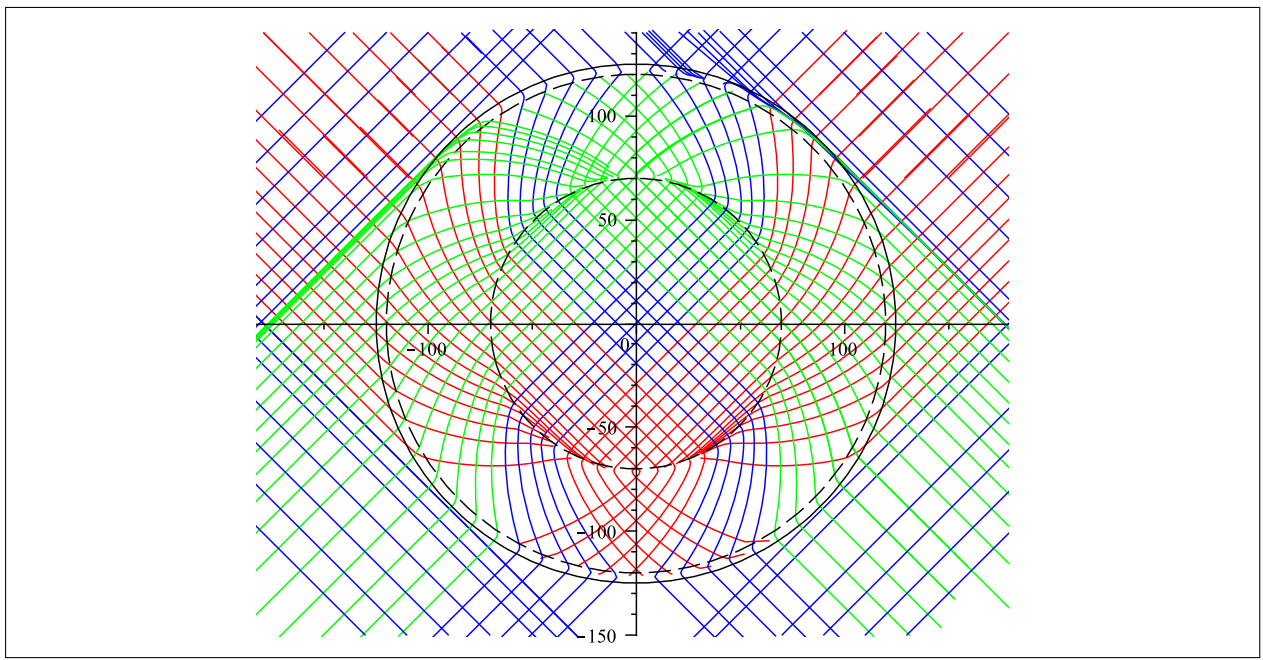}

\caption{This is a computer-generated spacetime diagram of how rays of light will travel through a one another of the TARDIS geometry. The black circles  represent the edges of the bubble. Rays of light end up getting strongly diffracted whenever they encounter the edges of the bubble.\label{hello}}
\end{figure*}

To consider the matter in more detail, we have plotted the trajectory of light rays as they cross the spacetime (See Fig. \ref{hello}). Most of the curvature is located at the walls of the bubble, so we see  most of the bending of the light occurring near  the walls. This type of diagram is helpful because it shows us which parts of the spacetime will be visible to other parts, and also because it details the orientation of lightcones.  Based on Fig. \ref{hello}, Amy may also be able to (occasionally) catch glimpses of herself in the past and future.

We should add that travelling inside the TARDIS bubble comes at a price. Travellers within the bubble must constantly accelerate to remain in place. Furthermore, the narrower the circle traced in spacetime, the larger the acceleration required. It seems that larger jaunts through space and time are easier on the body than short hops.

 \begin{figure} \caption{To build a TARDIS we require matter which has strange and unphysical properties. These graphs demonstrate that this matter must have negative pressure.  }
                             \label{fig:y0t0}
          
              \begin{subfigure}[b]{0.45\textwidth}
                             \includegraphics[trim=14mm 5mm 6mm 4mm, clip, width=8cm]{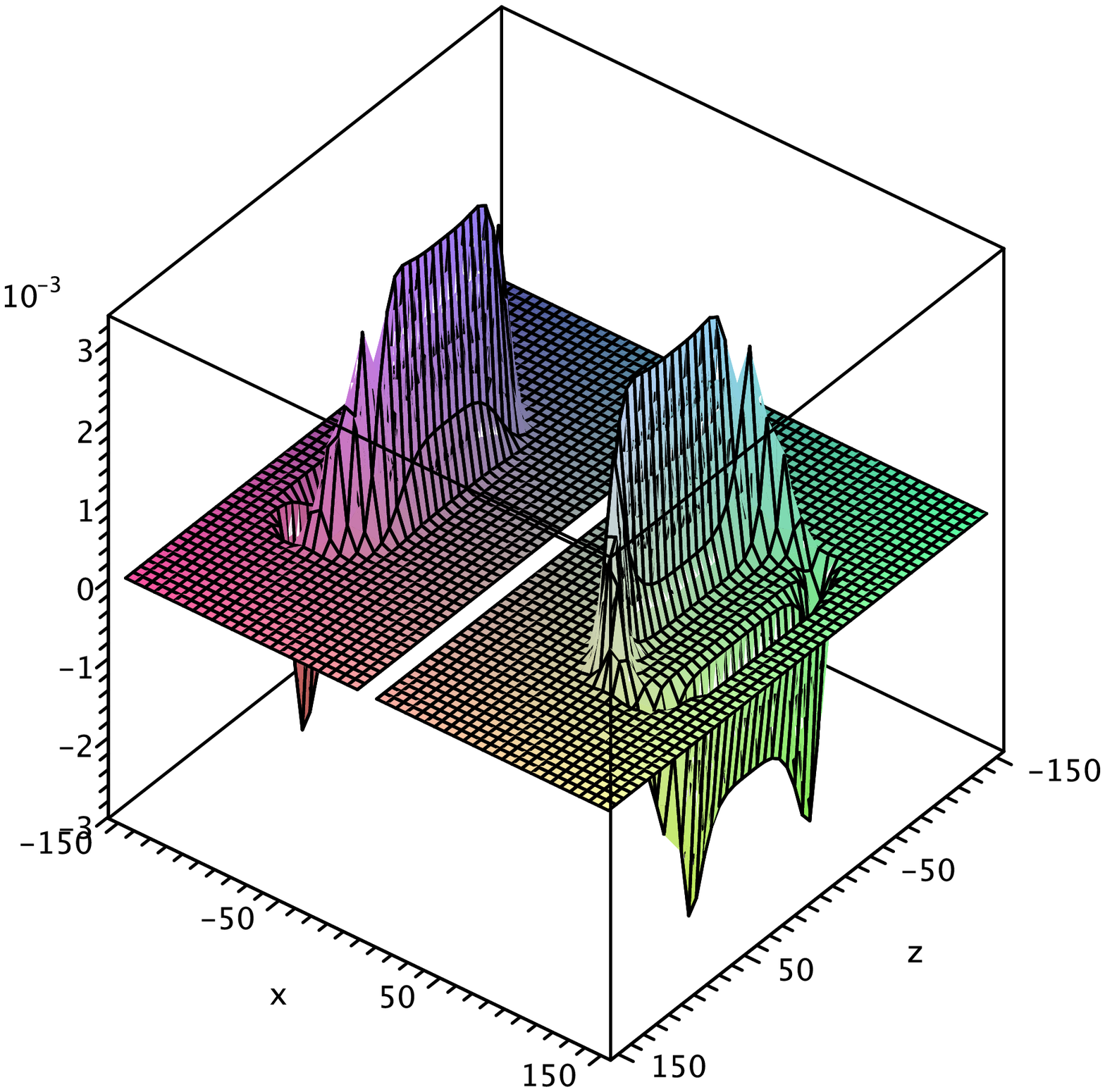}
                             \caption{Pressure in the $z$ direction along the cross-section $y=0,\, t=0$}
                             \label{fig:p3}
                   \end{subfigure}  
              \begin{subfigure}[b]{0.45\textwidth}
                             \includegraphics[trim=14mm 5mm 6mm 4mm, clip, width=8cm]{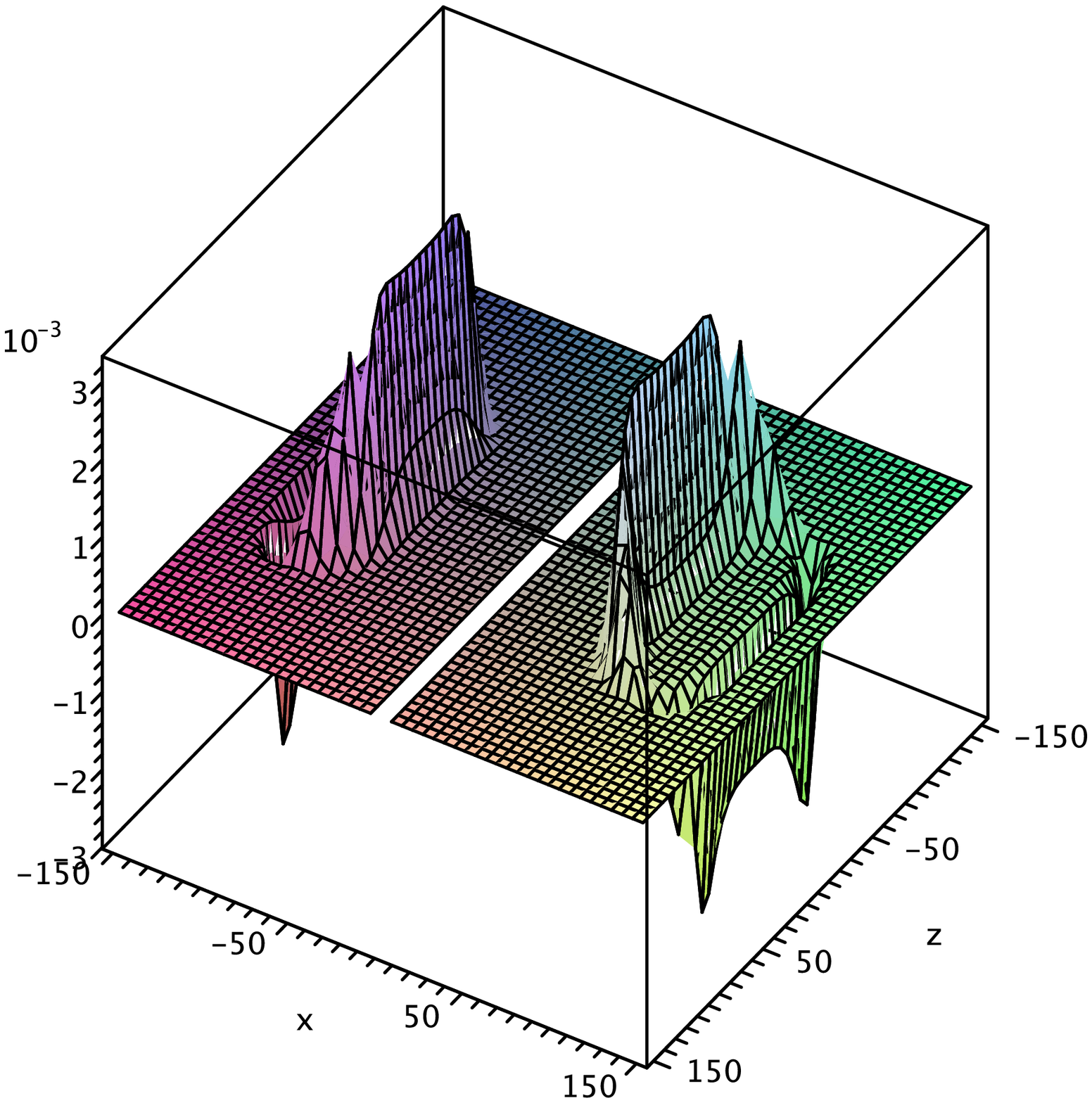}
                             \caption{Pressure in the $y$ direction along the cross-section $y=0,\, t=0$}
                             \label{fig:p4}
                   \end{subfigure}
                                
 \end{figure}
Could a TARDIS time machine ever be constructed? As we have discussed, we use the Einstein Equation to determine the matter's character and distribution (see Fig. \ref{fig:y0t0}). Unfortunately, just like the Alcubierre Warp Drive, generating the TARDIS geometry would require exotic matter (violating the classical energy conditions). This matter  would be gravitationally repulsive and would need to move faster than light.

\subsection*{Travelling Through Time and Space}

\begin{figure}
\includegraphics[trim=70mm 42mm 30mm 18mm, clip, width=12cm]{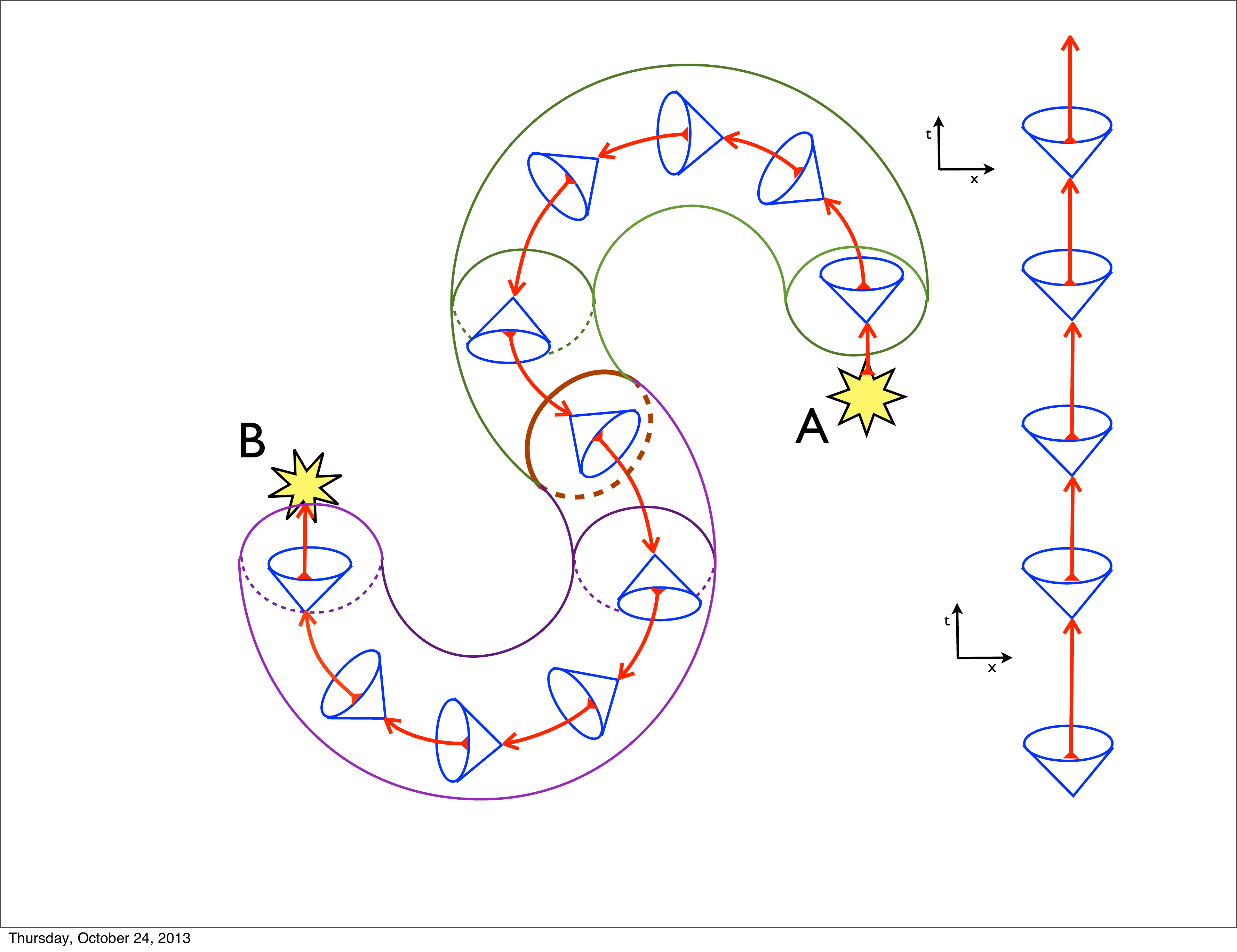}

\caption{Sections of different TARDIS geometries (green and purple) can be spliced together (along the orange boundary) in order to put any point (A) in the immediate causal past of any other point (B) \label{waterslideTAR}.}

\end{figure}

\begin{figure}
\includegraphics[trim=55mm 80mm 95mm 20mm, clip, width=10cm]{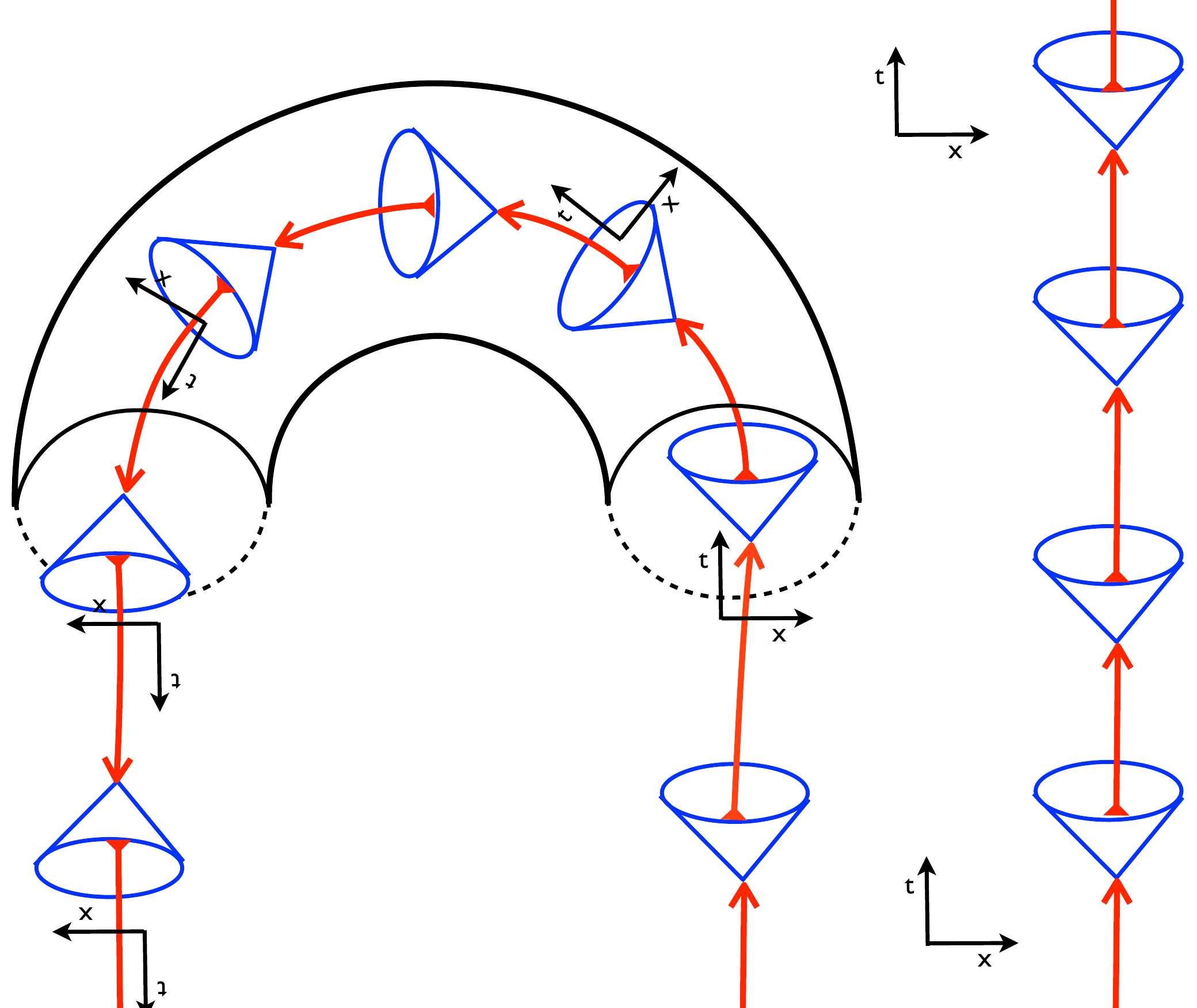}

\caption{An object which enters a TARDIS half-pipe will emerge the mirror image of itself, moving backwards in time. This could be considered antimatter. \label{antimatter}}

\end{figure}
Even if we could construct the TARDIS geometry, in the form we have presented it would be of limited value to explorers. Who wants to go in a circle? Circles are boring.

The true power of the TARDIS geometry is unlocked through the use of another clever technique: the spacetime equivalent of scissors and glue. Segments of different spacetime geometries can be mathematically cut up and then spliced together to create novel composite geometries.

Suppose I find myself at one point in spacetime (A), and I want to explore another point (B) which lies outside my lightcone. By cutting and splicing a series of TARDIS bubble  sections together, I could travel to the recent past of point (B) (See Fig. \ref{waterslideTAR}).

An alternative way to use the aforementioned  technique involves cutting a TARDIS geometry in half, resulting in an open-ended U-shape (Fig. \ref{antimatter}). Travellers who enter the geometry on one side would exit the other moving backwards in time with their right and left directions switched.

This is notable because the physical laws of our universe are time-symmetric in a specific way (called CPT symmetry: Charge, Parity, Time). If you take a particle and then switch the direction of the arrow of time, and then switch left-and-right, the equations which govern it will look the same as if we had switched the sign of the electric charge. Simply put: antimatter is just regular matter moving backwards in time\footnote{If you think that this idea is fun, look up the \emph{one-electron universe hypothesis}}.

Thus, it is possible that a traveller who has passed through a half-TARDIS  geometry will will end up (by virtue of it moving backwards in time) being made of antimatter. Alternatively, from the traveller's perspective,  she will emerge  into a universe  made of antimatter\footnote{In ``The Three Doctors'' the ancient Time Lord villain Omega converts the Second and Third Doctors, as well as their companions, into antimatter. Perhaps his mastery of TARDIS geometries allowed this to occur.}.  Of course, if you were standing outside the TARDIS,   you would see two bubble chambers, with antimatter entering one end, and regular matter entering the other, and then you would see the two chambers move together and disappear\footnote{The half-TARDIS will then resemble the process of \emph{Annihilation} from quantum field theory.}.

\section*{Summary}

We began this white paper with a brief explanation of Einstein's theory of curved spacetime: General Relativity. We discussed the lightcone as a means of illustrating the causal structure and curvature of a spacetime geometry. We then explained other proposed methods for using curved spacetime geometry to travel backwards in time. 

We introduced the basic properties of our TARDIS spacetime geometry. TARDIS is an acronym which stands for Traversable Achronal Retrograde Domains In Spacetime. Our TARDIS is a hollow bubble which allows its contents to travel along a closed, circular path in space and time. Travellers within the bubble will feel a persistent acceleration. People outside the TARDIS will see \emph{two} bubble chambers:  within one, time is evolving forwards; within the other, time evolves backwards. 

Building the TARDIS bubble geometry would  require an exotic type of matter which violates the classical energy conditions. We could not generate the required spacetime curvature using normal matter.

Finally, we discussed other ways the TARDIS geometry could be used. By cutting up different TARDIS tubes and joining them together, the new geometry could transport a traveller  between any two points in space and time. If a TARDIS geometry is cut in half, a traveller who enters one end may emerge into a universe made of antimatter. 

\section*{About the Authors}

Dr. Ben Tippett is an expert in Einstein's theory of general relativity, and he studies black holes. He has written several papers on fiction science, including his \emph{Unified Theory of Superman's Powers}, and a paper describing the gravitational warping of spacetime generated by Cthulhu's tomb city of R'lyeh. He is currently a sessional lecturer at the University of British Columbia Okanagan, where he teaches Mathematics. You can follow him on twitter at @bnprime, and you can ask him questions on tumblr at bnprime.tumblr.com .

Dr. Dave Tsang is an astrophysicist who studies planets, black holes and neutron stars. He loves Doctor Who, drawing silly pictures and thinking about physics. He is currently a postdoctoral researcher in Physics at McGill University. You can follow him on twitter at @DrDa5id.

Ben and Dave collaborate on \emph{The Titanium Physicists Podcast}\footnote{www.titaniumphysics.com} where expert physicists explain complicated physics to regular people using crazy analogies and explanations involving bees and motorcycles. If you would like to learn more about quantum mechanics, spacetime curvature, and black holes, why not listen while you drive to work? The Titanium Physicists Podcast is available on Stitcher, iTunes, the Apple Podcast App, and other places podcasts are found.

\bibliography{WhitePaper}
\bibliographystyle{plain}

\end{document}